\documentclass[prd ]{revtex4}%
\usepackage{caption}
\usepackage{amsmath}
\usepackage{graphicx}
\usepackage{epsfig}
\usepackage{dcolumn}
\usepackage{bm}
\usepackage{slashed}
\usepackage{amsfonts}
\usepackage{amssymb}%
\providecommand{\U}[1]{\protect\rule{.1in}{.1in}}
\begin{document}
\title{Is a generalized NJL model the effective action of massless QCD? }
\author{Alejandro Cabo Montes de Oca,}
\affiliation{Departamento de F\'{\i}sica Te\'{o}rica, Instituto de
Cibern\'{e}tica,Matem\'atica y F\'isica, Calle E, no. 309, Vedado, La Habana,
Cuba. }

\begin{abstract}
\noindent A local and gauge invariant alternative version of QCD for massive
fermions introduced in previous works, is considered here to just propose a
theory which includes Nambu-Jona-Lasinio (NJL) terms in its defining action in
a renormalizable form. The Lagrangian includes a special kind of new vertices
which at first sight, look as breaking power counting renormalizability.
However, these terms also modify the quark propagators, to become more
decreasing that the Dirac propagator at large momenta, indicating that the
theory is renormalizable. Therefore, it follows the surprising conclusion that
the added NJL four fermions terms does not break renormalizability. The
approach, can also be interpreted as a slightly generalized renormalization
procedure for massless QCD, which seems able to incorporate the mass
generating properties for the quarks of the NJL model, in a renormalizable
way. The structure of the free propagator, given by the substraction between a
massive and a massless Dirac one in the Lee-Wick form, also suggests that the
theory retains unitarity, if the radiative corrections make the massless
quarks become non propagating. The appearance of finite masses in the theory is justified
by the fact  that the new action terms break chiral invariance.  The scheme looks as being
able to implement the Fritzsch Democratic Symmetry breaking approach to quark mass hierarchy.
It seems also possible to further link the theory with  the SM  after employing the
Zimmermann's coupling constant reduction scheme in a similar way as the Top quark
condensation model had been approximately reformulated as a Higgs field one.

\end{abstract}
\maketitle

Determining the origin of the wide range of values spanned by the quark
masses, and more generally, the structure of the lepton and quark mass
spectrum, is one of the central problems of High Energy Physics. We have
considered a previous research activity associated to this question. It  was
motivated by the suspicion about that the large degeneration of the
non-interacting massless QCD vacuum (the state which is employed for the
construction of the standard Feynman rules of PQCD) in combination with the
strong forces carried by the QCD fields, could be able to generate a large
dimensional transmutation effect. This effect in turns, could then imply  the
generation of  quark and gluon condensates and  masses. The investigation of
quark and gluon condensation effects had been widely considered in the
literature \cite{1,2,3,4,5,6,7,8}. Our previous works on the
theme appeared in references \cite{9,10,11,12,13,14,15,16,17}. \  Assumed
that the idea in them is correct,  the following picture could arise. A sort of
Top condensate model might be the effective action for massless QCD. In this
case, a  Top quark condensate,  arising within the same inner context of the
SM,  could play the role of the Higgs field. Thus, the SM could be
\textquotedblleft closed\textquotedblright\ by generating all the masses
within its proper framework. We  could imagine this effect to occurs as
follows (see the figure 1 for illustrating  the argue): In a first step, the
six quarks could get their masses thanks to a flavor symmetry breaking
determined by the quark and gluon condensates. Afterwards, the electron, muon
and tau leptons, would receive their intermediate masses thanks to radiative
corrections mediated by the mid strength electromagnetic interactions with
quarks. \ Finally, the only weak interacting character of the three neutrinos
with all the particles, could determine their even smaller mass values.
\begin{figure}[h]
\begin{centering}
\center
\includegraphics[scale=0.50]{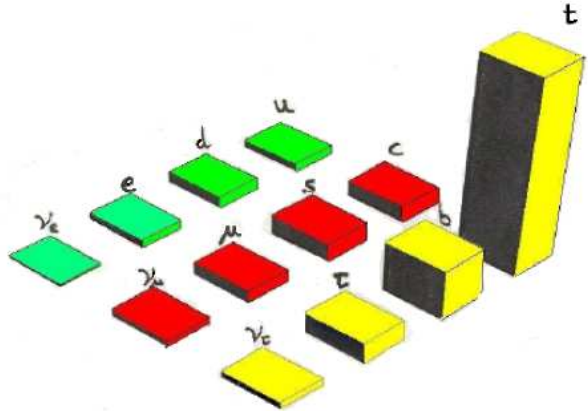}
\par\end{centering}
\caption{{}A qualitative illustration (the mass ratios are not correct ones)
of the particle mass spectrum. Note the  similar mass increasing behavior with
the Family number for quarks and leptons. In addition, the reduction of the
mass of the electron, muon and tao leptons with respect to the quarks, could
be suspected to be due to the only electromagnetic strongest  interaction of
these particles. Further, the even smaller mass scales of the three neutrinos
could be originated in the only weakly interacting character of them.   }%
\label{fig1}%
\end{figure}

Its is clear, that such a picture will need to satisfy  the many experimental
constraints imposed by the strong experimental evidence about the validity of
SM model which had came from the LHC. However, we have no $apriori$
sufficiently appealing reasons to discard the already obtained indications
about the possibility for the idea to be correct.

Before starting to present the proposal of the model, let us briefly review
the previous results which motivated it. In Ref. \cite{10}, with the use of a
BCS squeezed state like vacuum state (formed with nearly zero momenta gluons
and ghost particles) modified Feynman rules for massless QCD were derived. The
case of gluon condensation in the absence of quark pair condensation was
initially considered. Then, a  proper selection of the parameters allowed to
derive an addition to the gluon free propagator: a Dirac's Delta function
centered at zero momenta multiplied by the metric tensor. Such a term were
before discussed by Munczek and Nemirovsky in \cite{2}. Before, in reference
\cite{9} it was simply proposed this modification and used to argue that it
predicts a non vanishing value of the gluon condensate in the first
corrections. The physical state and zero ghost number conditions were also
imposed to fix the parameters of the squeezed vacuum. Then, the results
obtained for gluons motivated the idea of also considering the quarks as
massless and to search for the possibility to generate their masses
dynamically, thanks to the condensation of quark pairs. For this purpose the
BCS like initial state was generalized in reference \cite{11} to include the
quark pair condensates in massless QCD. In this case, in a similar way as for
gluons, the quark propagators simply were modified again by the addition of
a term being the product of a Dirac's Delta function at zero momentum and the
spinor identity matrix. Next, in Refs. \cite{11,12} the main conclusion obtained
from this starting approach followed from a simple discussion of the Dyson
equation for quarks. It was considered by taking the quark self-energy in its
lowest order in the power expansion in the condensate parameters. The
coefficient of the zero momentum Delta function was fixed to reproduce the
estimate of the gluonic Lagrangian mean value, following from the sum rules
approaches. After that,  the solution of the Dyson equation was able
to predict the \textquotedblleft constituent\textquotedblright\ values of 1/3
of the nucleon mass for the light quarks.  The initial approach was also
further studied  in \cite{13,14}  in order to define a regularization scheme for
 eliminating the singularities which could appear in the Feynman diagram
expansion due to the Delta functions at zero momentum entering in the modified
propagators.

However, due to the unusual characteristics of the approach, we decided in reference
\cite{15,16} to also investigate the possibility of re-expressing the condensation
effects in the modified propagators as equivalent vertices in the Lagrangian.
The result of this work was  the central step in leading to the model here
presented. \ It was obtained  that the condensate effects introduced in
massless QCD by employing a squeezed state as modified vacuum, were equivalent
to the addition of  a new four legs vertex term in the Lagrangian, including
 two gluon and two quark lines. But, the \ vertex was a non local one including a
zero momentum delta function. \ The obtained vertex structure clearly was not
a fully gauge invariant one,  but this drawback can be understood as due \ to the
simple non gauge invariant form employed \ for the squeezed state vacuum. \ However,
the curious structure directly led to the idea of constructing a local and
gauge invariant form of the theory including a similar kind of two gluon and
two quark vertices, presumably incorporating the gluon and quark
condensates in a  gauge invariant and local form. This modification was
presented in reference   \cite{17}, where it was also argued that the new terms of
the action  do not break the power counting renormalizability  of massless
QCD. \ This observation,  then led to a surprising conclusion also exposed in
\ reference  \cite{17}, that the Nambu-Jona-Lasinio (NJL) four fermion vertices
turn out to become renormalizable counterterms of the considered Lagrangian.
\ \ \ The resulting theory included an additional  set of  six fermion fields
showing \ a negative metric. However, the \ modified \ quark propagator also
showed a rapidly\ decaying momentum dependence \ thanks to its Lee-Wick
structure \cite{leewick1} which suggests that the negative metric states \ could
result to be non propagating thanks  to the radiative corrections  \cite{leewick2}.
\ The mentioned properties  opened  the opportunity that the proposed  models
can show the mass generation properties which the NJL models exhibit.  The
study of this possibility will be a main objective of  the extensions of the this  work.

\ Here, the previous \ results will be employed to motivate and  propose \ a
local and gauge invariant form of QCD including  the special two gluons and
two quark vertices and NJL four fermion terms and  being \ power counting
renormalizable.  In section II, firstly we present the argue which allowed to
represent the quark and gluon condensation effects in the previous discussion
as a two-gluon-two-quark vertices in the massless QCD action. This section is a
review of the results in \cite{16}. \ Section III continues by resuming  the
construction done in \cite{17}  of  a local and  gauge invariant Lagrangian for
massless QCD,   by promoting the non local and gauge non-invariant action
derived in the previous section to be local and gauge invariant. \ The given
action expression included both gluon and quark condensate effects. However,
in the next Section IV,  as it was done in reference \ \cite{17}, the gluon
condensation effects, which are assumed to contribute more importantly to low
energy effects and confinement, are initially disregarded for to be considered
separately. \ In this case the  action following \ was simply the massless QCD
one \  plus six additional terms, one for each flavor,  of similar vertices
formed by products of two quark  and two covariant derivatives. Further,
Section V \ defines a final form for  the here  proposed model  which in
addition includes the  NJL four quark vertices, which are  now allowed by the power
counting renormalizability. There is also defined the specific notations to be
employed in the extensions of this work.  Section VI simply presents a qualitative argue
suggesting that the proposed model can result to be unitary thanks to its similarities
with the Lee-Wick theories.  Next, Section VII also briefly exposes the possibilities
that the discussed perturbative expansion can generate a quark mass hierarchy though
a flavor symmetry breaking effect.   Finally the Summary resume the content of the work and
pose some remarks on its possible implications.

\section{From modified propagators to a  modified Lagrangian}

In this section we will describe how in reference \cite{16},  the gluon and
quark condensate parameters appearing in the propagators, were "shifted" to
appear as new couplings in a modified action for massless QCD. The Feynman
diagram expansion of the modified theory was defined by the generating
functional of Green functions having the form:%
\begin{align}
Z[j,\eta,\overline{\eta},\xi,\overline{\xi}] &  =\frac{I[j,\eta,\overline
{\eta},\xi,\overline{\xi}]}{I[0,0,0,0]},\nonumber\\
I[j,\eta,\overline{\eta},\xi,\overline{\xi}] &  =\exp(V^{int}[\frac{\delta
}{\delta j},\frac{\delta}{\delta\overline{\eta}},\frac{\delta}{-\delta\eta
},\frac{\delta}{\delta\overline{\xi}}\frac{\delta}{-\delta\xi}]){\small \times
}\label{Wick}\\
&  \exp(\int\frac{dk}{(2\pi)^{D}}j(-k)\frac{1}{2}D(k)j(k)){\small \times
}\nonumber\\
&  \exp(\sum_{f}\int\frac{dk}{(2\pi)^{D}}\overline{\eta}_{f}(-k)G_{f}%
(k)\eta_{f}(k)){\small \times}\nonumber\\
&  \exp(\int\frac{dk}{(2\pi)^{D}}\overline{\xi}(-k)G_{gh}(k)\xi(k)){\small ,}%
\nonumber\\
f &  =1,2,...6.
\end{align}
This functionals is associated to the action $S_{g}$ depending on
the gauge interaction coupling $g$,  and \ $S_{0}$, which is $S_{g}$ for $g=0$, defines the free
action. The action  $S_{g}$ and the vertex part \ Lagrangian $V^{int}$ are
then defined in terms of the six quark $\Psi_{f}$ ($f=1,...,6$), gluons $A$ and
ghost $\chi$ fields in the usual massless QCD form
\begin{align}
S_{g} &  =\int dx(-\frac{1}{4}F_{\mu\nu}^{a}F_{\mu\nu}^{a}-\frac{1}{2\alpha
}\partial_{\mu}A_{\mu}^{a}\partial_{\nu}A_{\nu}^{a}+\overline{c}%
^{a}\overleftarrow{\partial}_{\mu}D_{\mu}^{ab}c^{b}-\sum_{f}\overline{\Psi
}_{f}^{i}\text{ }i\gamma_{\mu}D_{\mu}^{ij}\Psi_{f}^{j}),\text{ \ }\\
V^{int} &  =S_{g}-S_{0},\label{vint}%
\end{align}
where the field intensity and covariant derivatives follow the Euclidean
conventions
\begin{align}
F_{\mu\nu}^{a} &  =\partial_{\mu}A_{\nu}^{a}-\partial_{\nu}A_{\mu}%
^{a}-gf^{abc}A_{\mu}^{b}A_{\nu}^{c},\nonumber\\
D_{\mu}^{ij} &  =\partial_{\mu}\delta^{ij}+ig\text{ }A_{\mu}^{a}T_{a}%
^{ij},\ \ \ \ D_{\mu}^{ab}=\partial_{\mu}\delta^{ab}+gf^{abc}\text{ }A_{\mu
}^{c},\\
\{\gamma_{\mu},\gamma_{\nu}\} &  =-2\delta_{\mu\nu},\ \ \ \ [T_{a}%
T_{b}]=if^{abc}T_{c}.\nonumber
\end{align}

The $D$, $G_{f}$ and $G_{gh}$ are the mentioned in the Introduction modified
gluon, quark and ghost propagators including condensate effects:%

\begin{align}
D_{\mu\nu}^{ab}(k) &  =\delta^{ab}(\frac{1}{k^{2}}(\delta_{\mu\nu}%
-(1-\alpha)\frac{k_{\mu}k_{\nu}}{k^{2}})\theta_{N}(k)+C_{g}\delta^{D}%
(k)\delta_{\mu\nu}),\nonumber\\
G_{f}^{ij}(k) &  =\delta^{ij}{\large (}\frac{\theta_{N}(k)}{m+\gamma_{\mu
}k_{\mu}}+C_{f}\delta^{D}(k)I{\large )},\nonumber\\
G_{gh}^{ab}(k) &  =\delta^{ab}\frac{\theta_{N}(k)}{k^{2}}.\label{propagators}%
\end{align}
Note, that the propagators are basically the same Feynman's ones of \ PQCD,
after adding Dirac delta functions at zero momentum (\cite{2}) that represent
the gluon and quark condensates. They also
include the  Heaviside functions, introduced by Nakanishi to solve difficulties
in the quantization of the free gauge theory \cite{nakanishi}. These functions
make the Feynman contribution to the propagator vanish in a small neighborhood
of the zero value of the momentum. As mentioned before, their consideration
allowed in reference \cite{13} to get rid of
various singular contributions to the Feynman expansion that could have been
appeared, due to the distributional character of the Dirac delta function
terms in the propagators.

Then, in reference \cite{16}, for the purpose of \textquotedblleft shifting
\textquotedblright\ the gluon and quark condensate parameters to appear in
vertex terms within a modified action for massless QCD,  it was employed the
property that the "condensates" corrections to the propagators define
quadratic forms in the zero momentum component of the spatial Fourier
transforms of the sources. Then, the exponential of those terms in the
generating functional $Z$ were expressed as Gaussian integrals over auxiliary
vector and fermion parameters, not functions. This change transformed in
linear ones the dependence of the Feynman integrands defining $Z$ on the
sources which  are contracted with the condensates dependent part of the
propagators. Then, after evaluating the commutator of an exponential having an
argument being linear in the sources, with the exponential of the vertex part
$V_{int}$ as expressed in terms of the functional derivatives over the
sources, the generating functional $Z$ was written in the form%
\begin{align}
Z[j,\eta,\overline{\eta},\xi,\overline{\xi}] &  =\frac{1}{\mathcal{N}}\int\int
d\alpha d\overline{\chi}d\chi\exp[-\sum_{f}\overline{\chi}_{f,\text{ }r}%
^{i}\chi_{f,\text{ }r}^{i}-\frac{\alpha_{\mu}^{a}\alpha_{\mu}^{a}}%
{2}]\nonumber\\
&  \exp{\large [}{\small S}_{g}^{\ast}{\small [\frac{\delta}{\delta j}%
,\frac{\delta}{\delta\overline{\eta}},\frac{\delta}{-\delta\eta},\frac{\delta
}{\delta\overline{\xi}},\frac{\delta}{-\delta\xi},\alpha,\chi,\overline
{\chi}]{\large ]}\times}\label{Zmod}\\
&  \int\mathcal{D}[A,\overline{\Psi},\Psi,\overline{c},c]\exp{\large [}\int
dx\text{ }{\large (}{\small j(x)A(x)+\sum_{f}{\large (}\overline{\eta}%
_{f}(x)\Psi_{f}(x)+\overline{\Psi}_{f}(x)\eta_{f}(x){\large )}{\Large )}%
}{\large ]}\nonumber\\
&  =\frac{1}{\mathcal{N}}\int\int d\alpha d\overline{\chi}d\chi\exp
{\large [}-\sum_{f}\overline{\chi}_{f,\text{ }r}^{i}\chi_{f,\text{ }r}%
^{i}-\frac{\alpha_{\mu}^{a}\alpha_{\mu}^{a}}{2}{\large ]}\times\nonumber\\
&  \int\mathcal{D}[A,\overline{\Psi},\Psi,\overline{c},c]\exp{\large [}\int
dx\text{ }{\large (}S_{g}^{\ast}{\small [}A{\small ,\Psi,}\overline{\Psi
},c,\overline{c},\alpha,\chi,\overline{\chi}{\small ]+}\nonumber\\
&  {\small j(x)A(x)+\sum_{f}{\large (}\overline{\eta}_{f}(x)\Psi
_{f}(x)+\overline{\Psi}_{f}(x)\eta_{f}(x){\large ) + \small \overline
{{\small \xi}}(x)c(x)+\overline{c}(x)\xi(x)
)}{\Large {\large ]}.}%
}\nonumber
\end{align}
where the changed action $S_{g}^{\ast}$ had  the form%
\begin{align}
S_{g}^{\ast} &  =S_{g}^{\ast}[A{\small ,}\Psi{\small ,}\overline{\Psi
}{\small ,}c,\overline{c},{\small \alpha,\chi,\overline{\chi}]}\nonumber\\
&  =\int dx{\large [}{\Large -}\frac{1}{4}F_{\mu\nu}^{a}(A+{\small (}%
\frac{2C_{g}}{(2\pi)^{D}}{\small )}^{\frac{1}{2}}\alpha_{\mu}^{a})F_{\mu\nu
}^{a}(A+{\small (}\frac{2C_{g}}{(2\pi)^{D}}{\small )}^{\frac{1}{2}}\alpha
_{\mu}^{a})\nonumber\\
&  -\frac{1}{2\alpha}\partial_{\mu}A_{\mu}^{a}\partial_{\nu}A_{\nu}%
^{a}-\overline{c}^{a}\partial_{\mu}D_{\mu}^{ab}(A+{\small (}\frac{2C_{g}%
}{(2\pi)^{D}}{\small )}^{\frac{1}{2}}\alpha_{\mu}^{a})c^{b})\nonumber\\
&  -\sum_{f}{\Large (}\overline{\Psi}_{f}^{i}\text{ }i\gamma_{\mu}D_{\mu}%
^{ij}(A+{\small (}\frac{2C_{g}}{(2\pi)^{D}}{\small )}^{\frac{1}{2}}\alpha
_{\mu}^{a})\Psi_{f}^{j}+\overline{\Psi}_{f}^{i}\text{ }i\gamma_{\mu}D_{\mu
}^{ij}(A)\chi_{f}^{j}{\small (}\frac{C_{f}}{(2\pi)^{D}}{\small )}^{\frac{1}%
{2}}+{\small (}\frac{C_{f}}{(2\pi)^{D}}{\small )}^{\frac{1}{2}}\overline{\chi
}_{f}^{i}\text{ }i\gamma_{\mu}D_{\mu}^{ij}(A)\Psi_{f}^{j}\nonumber\\
&  +{\small (}\frac{C_{f}}{(2\pi)^{D}}{\small )(}\frac{2C_{g}}{(2\pi)^{D}%
}{\small )}^{\frac{1}{2}}\overline{\chi}_{f}^{i}\text{ }i\gamma_{\mu}\text{
}ig\alpha_{\mu}^{a}T_{a}^{ij}\chi_{f}^{j}{\large )]}{\Large .}%
\end{align}

This was the expression which directly suggested the local form for QCD
proposed in reference \cite{17}. \ Thus, let us resume below the elements  from
the above relations leading to the proposal. Firstly, note that the auxiliary
boson and fermion  parameters ($\alpha_{\mu}^{a}$ and $\chi_{f}^{j}$)
appearing thanks to the  Gaussian integral representations of the condensate
dependent sources terms, are constants independent of the space-time
coordinates. As already pointed out before, this property was a direct
consequence of the simple perturbative modifications of the free QCD vacuum
employed to connect the interaction in the construction of the Wick expansion.
Therefore, all the space time derivatives of these parameters in the above
formula, in fact vanish, which makes that the expression can not be explicitly
seen as corresponding to a quantized gauge invariant theory. This represents a
conceptual limitation of the version of the modified massless QCD before
considered. However, the above written form of the action suggests a direct
possibility to overcome this problem. It  consists in simply modifying this
action to become gauge invariant and local as suggested by the various   gauge
invariant constitutive elements entering its structure. The reasonable
character of this idea to modify the action, comes from the fact that the
obtained generating functional was derived, by considering very particular
forms of the initial modified free vacuum states incorporating gluon and quark
condensates. Therefore, it can be conceived that the gradual connection of the
interaction (or a better constructed initial state) could eventually lead to
the below simply to be proposed gauge invariant \ form of the action.
In the next section, this construction will be reviewed.

\section{The local form of the model}

Then, in this section we will resume the proposal done in reference
\cite{17} of a local and gauge invariant QCD Lagrangian including gluon
and quark condensate effects in a way resembling the former study. The
construction started by \textquotedblleft promoting\textquotedblright\ the
constant space-time independent condensate parameters to be full space-time
dependent functions. Further, the new action (without considering the
Fadeev-Popov gauge fixing terms) was taken as the same one \ as in the
previous modified QCD, but  in which now the new gauge and fermion fields
$\alpha,\overline{\chi},\chi$ fields transform in a homogeneous way under the
same gauge transformation leaving invariant the massless QCD action. The
generating functional expression resulted in  the form %

\begin{align}
Z[j,\eta,\overline{\eta},\xi,\overline{\xi}] &  =\frac{1}{\mathcal{N}}\int%
\int\mathcal{D}[\alpha,\overline{\chi},\chi]\exp[-\sum_{f}\overline{\chi
}_{f,\text{ }r}^{i}(x)\chi_{f,\text{ }r}^{i}(x)-\frac{\alpha_{\mu}%
^{a}(x)\alpha_{\mu}^{a}(x)}{2}]\times\nonumber\\
&  \int\mathcal{D}[A,\overline{\Psi},\Psi,\overline{c},c]\exp{\large [}\int
dx\text{ }{\large (}S_{g}^{\ast}{\small [}A{\small ,\Psi,}\overline{\Psi
},c,\overline{c},\alpha,\chi,\overline{\chi}{\small ]+}\nonumber\\
&  {\small j(x)A(x)+}\sum_{f}{\large (}{\small \overline{\eta}}_{f}%
{\small (x)\Psi}_{f}{\small (x)+\overline{\Psi}}_{f}{\small (x)\eta}%
_{f}{\small (x){\large )+ \small \overline
{{\small \xi}}(x)c(x)+\overline{c}(x)\xi(x)
)]}},\label{gluoquark}%
\end{align}
where now, \ the local $S_{g}^{\ast}$ action is defined as%

\begin{align}
S_{g}^{\ast}  &  =S_{g}^{\ast}[A{\small ,}\Psi{\small ,}\overline{\Psi
}{\small ,}c,\overline{c},{\small \alpha,\chi,\overline{\chi}]}\nonumber\\
&  =\int dx{\Large [}-\frac{1}{4}F_{\mu\nu}^{a}(A+{\small (}\frac{2C_{g}%
}{(2\pi)^{D}}{\small )}^{\frac{1}{2}}\alpha_{\mu}^{a})F_{\mu\nu}%
^{a}(A+{\small (}\frac{2C_{g}}{(2\pi)^{D}}{\small )}^{\frac{1}{2}}\alpha_{\mu
}^{a})\nonumber\\
&  -\frac{1}{2\alpha}\partial_{\mu}A_{\mu}^{a}\partial_{\nu}A_{\nu}%
^{a}-\overline{c}^{a}\partial_{\mu}D_{\mu}^{ab}(A)\text{ }c^{b})\nonumber\\
&  -\sum_{f}\overline{\Psi}_{f}^{i}\text{ }i\gamma_{\mu}D_{\mu}^{ij}%
(A+{\small (}\frac{2C_{g}}{(2\pi)^{D}}{\small )}^{\frac{1}{2}}\alpha_{\mu}%
^{a})\text{ }\Psi_{f}^{j}\nonumber\\
&  -\sum_{f}\overline{\Psi}_{f}^{i}\text{ }i\gamma_{\mu}D_{\mu}^{ij}%
(A)\chi_{f}^{j}\text{ }{\small (}\frac{C_{f}}{(2\pi)^{D}}{\small )}^{\frac
{1}{2}}-\sum_{f}{\small (}\frac{C_{f}}{(2\pi)^{D}}{\small )}^{\frac{1}{2}%
}\overline{\chi}_{f}^{i}\text{ }i\gamma_{\mu}D_{\mu}^{ij}(A)\Psi_{f}%
^{j}\nonumber\\
&  -\sum_{f}{\small (}\frac{C_{f}}{(2\pi)^{D}}{\small )(}\frac{2C_{g}}%
{(2\pi)^{D}}{\small )}^{\frac{1}{2}}\overline{\chi}_{f}^{i}\text{ }%
i\gamma_{\mu}\text{ }ig\alpha_{\mu}^{a}T_{a}^{ij}\chi_{f}^{j}{\Large ].}%
\end{align}

Note that now, the covariant derivatives over the new auxiliary fields
$\alpha,\overline{\chi},\chi$ include the spatial derivatives which were
absent in the former action. \ The bosonic spatially dependent field $\alpha
,$ entered in similar way that a gauge background field perturbation: that is,
transforming by the homogeneous part of the gauge transformation. It should be
also noticed that the action terms corresponding \ to the Fadeev-Popov gauge
fixation had been adopted in the usual Lorentz gauge. \ \ \ \ \

It can be remarked, that in the case of the vanishing quark condensate
parameters, we expect that the expansion can lead to interesting physical
consequences in the low energy region around $1$ GeV, were a similar
discussion in the previous non local expansion gave predictions \ for the
constituent quark masses of the light fermions and reproduced the Savvidy
chromomagnetic effective potential form, as a function of the gluon condensate
parameter \cite{11}. \ Further, an interesting circumstance is the fact that the
appearance of \ Gaussian integration over the new auxiliary fields directly
suggests a possible link with the so called "stochastic vacuum" vacuum
approach initiated by Dosch \cite{doetsch}.

However, in the rest of this article we will only study the case of the
vanishing gluon condensation parameter  $C_g$. It will be considered first \ by the
mentioned suspicion about the \ possibilities for quark mass generation in the
proposed  model.

\section{The vanishing gluon condensate limit}

As remarked before, in reference \cite{17} seeking to simplify the discussion in
a first stage, the limit of vanishing gluon condensate parameter was initially
assumed in (\ref{gluoquark}). Then, the $Z$ \ functional reduced to%
\begin{align}
Z[j,\eta,\overline{\eta},\xi,\overline{\xi}] &  =\frac{1}{\mathcal{N}}\int%
\int\mathcal{D}[\alpha,\overline{\chi},\chi]\exp{\large [}-\sum_{f}%
\overline{\chi}_{f,\text{ }r}^{i}(x)\chi_{f,\text{ }r}^{i}(x)-\frac
{\alpha_{\mu}^{a}(x)\alpha_{\mu}^{a}(x)}{2}{\large ]}\times\nonumber\\
&  \int\mathcal{D}[A,\overline{\Psi},\Psi,\overline{c},c]\exp{\large [}\int
dx\text{ }{\large (}S_{g}^{\ast}{\small [}A{\small ,\Psi,}\overline{\Psi
},c,\overline{c},\chi,\overline{\chi}{\small ]+}\nonumber\\
&  {\small j(x)A(x)+\sum_{f}{\large (}\overline{\eta}_{f}(x)\Psi
_{f}(x)+\overline{\Psi}_{f}(x)\eta_{f}(x){\large )}}+{\small \overline
{{\small \xi}}(x)c(x)+\overline{c}(x)\xi(x){\large ))]}},\label{Zquark}%
\end{align}
where the action adopts the simpler form%
\begin{align}
S_{g}^{\ast} &  =S_{g}^{\ast}[A{\small ,}\Psi{\small ,}\overline{\Psi
}{\small ,}c,\overline{c},{\small \chi,\overline{\chi}]}\nonumber\\
&  =\int dx\text{ }{\Large [-}\frac{1}{4}F_{\mu\nu}^{a}F_{\mu\nu}^{a}-\frac
{1}{2\alpha}\partial_{\mu}A_{\mu}^{a}\partial_{\nu}A_{\nu}^{a}-\overline
{c}^{a}\partial_{\mu}D_{\mu}^{ab}c^{b})\nonumber\\
&  -\sum_{f}{\Large (}\overline{\Psi}_{f}^{i}\text{ }i\gamma_{\mu}D_{\mu}%
^{ij}\Psi_{f}^{j}+\overline{\Psi}_{f}^{i}\text{ }i\gamma_{\mu}D_{\mu}^{ij}%
\chi_{f}^{j}\text{ }{\small (}\frac{C_{f}}{(2\pi)^{D}}{\small )}^{\frac{1}{2}%
}+{\small (}\frac{C_{f}}{(2\pi)^{D}}{\small )}^{\frac{1}{2}}\overline{\chi
}_{f}^{i}\text{ }i\gamma_{\mu}D_{\mu}^{ij}\Psi_{f}^{j}{\Large )]},
\end{align}
which is basically the massless QCD action plus two linear terms in the new
fermion fields $\chi$ and \ $\overline{\chi}$\ . \ Now, the Gaussian integral
over the auxiliary functions were evaluated by solving the Lagrange equations,
that after substituting the solutions for the fields in the expression for
$Z$ \ led to%

\begin{align}
Z &  =\frac{1}{\mathcal{N}}\int\mathcal{D}[A,\overline{\Psi},\Psi,\overline
{c},c,]\exp[{S[A,\overline{\Psi},\Psi,\overline{c},c]}],\label{Z01}\\
{S[A,\overline{\Psi},\Psi,\overline{c},c]} &  ={S}_{mqcd}{[A,\overline{\Psi
},\Psi,\overline{c},c]+\text{ }S^{q}[A,\overline{\Psi},\Psi]}\label{ac1}\\
S_{mqcd}[A,\overline{\Psi},\Psi,\overline{c},c] &  =\int dx(-\frac{1}{4}%
F_{\mu\nu}^{a}F_{\mu\nu}^{a}-\frac{1}{2\alpha}\partial_{\mu}A_{\mu}%
^{a}\partial_{\nu}A_{\nu}^{a}-\sum_{f}\overline{\Psi}_{f}^{i}\text{ }%
i\gamma_{\mu}D_{\mu}^{ij}\Psi_{f}^{j}-\overline{c}^{a}\partial_{\mu}D_{\mu
}^{ab}c^{b}),\label{ac2}%
\end{align}
where the action $S_{mqcd}$ is the usual one for massless QCD and as in
Ref. \cite{17}, new six action terms appeared, one for each quark flavor $f$.
\ They have the expressions%
\begin{equation}
S^{q}[A,\overline{\Psi},\Psi]=-\sum_{f}\frac{C_{f}}{(2\pi)^{D}}\int
dx\overline{\Psi}_{f}^{j}\text{ }i\gamma_{\mu}\overleftarrow{D}_{\mu}%
^{ji}\text{ }i\gamma_{\nu}D_{\nu}^{ik}\Psi_{f}^{k}.\label{ac3}%
\end{equation}

These new action terms are local and gauge invariant and most \ relevantly,
also they do not disturb power counting renormalizability,  because they also
 make the quark propagator to decrease with the square of the momentum. 

The four legs vertex is the local counterpart of the non local one appearing
in the previous expansion, which exactly represent massless QCD on a modified
free vacuum. The pure gluon and ghost one loop diagrams are the same as in
massless QCD because the gluon propagators is the same as in usual massless
QCD. \ The evaluated in \cite{17}  additional contributions to the one loop
polarization operator became transversal, thus satisfied the Ward identity
associated to the gauge invariance.

In ending this section, it should be remarked, that the action terms written
in (\ref{ac3}) are central elements in the present work. There is one of
such terms for each quark flavor and \ they introduce six dimensional
parameters which are their coupling constants. Also, they are precisely
corresponding to the quark condensate parameters of the former approach
discussed in \cite{16}, when the auxiliary fields are assumed as space-time
independent constants \ \ Their relevance comes from the fact that they do not
disturb the power counting renormalizability of massless QCD, thanks to the
fact that they modify the original Dirac's propagator to show a more rapidly
decreasing behavior at large momenta. This special property, after taking into
account the need of renormalizing the theory, with its central procedure of
\ determining the appropriate counterterms, opens the surprising opportunity
of \ consider  the standard Nambu-Jona-Lasinio action terms as appropriate
counterterms. This inclusion in the normal case is completely  excluded due to
slowly decreasing behavior of the Dirac's propagator. The described  outcomes
are employed in next section to propose a definite model for a local and gauge
invariant QCD including NJL terms in a \ renormalizable way.  \

\section{ A QCD model  including NJL terms in a renormalizable form}

\ \ The action of the here proposed model \ is written in  concrete the form:%

\begin{align}
S &  =\int dx\text{ }{\Large (}\,-\frac{1}{4}F_{\mu\nu}^{a}(x)F^{a\mu\nu
}(x)-\frac{1}{2\alpha}\partial_{\mu}A^{a\mu}(x)\partial_{\nu}A^{a\nu
}(x)+\overline{c}^{a}(x)\partial_{\mu}D^{ab\mu}c^{b}(x)-\nonumber\\
&  -\sum_{f}\overline{\Psi}_{f}^{i}(x)\text{ }i\gamma^{\mu}D_{\mu}^{ij}%
\Psi_{f}^{j}(x)-\sum_{f}\varkappa_{f}\,\overline{\Psi}_{f}^{j}\text{
}(x)\gamma_{\mu}\overleftarrow{D}^{ji\mu}\text{ }\gamma_{\nu}D^{ik\nu}\Psi
_{f}^{k}(x)\,{\LARGE )}+.\sum_{f,,f^{\prime}}\Lambda_{f_{1}\text{ }f_{2}%
f_{3}f_{4}}^{j_{1}j_{2}j_{3}j_{4}}\overline{\Psi}_{f_{1}}^{j_{1}}\text{
}(x)\Psi_{f_{2}}^{j_{2}}\text{ }(x)\overline{\Psi}_{f_{3}}^{j_{3}}\text{
}(x)\Psi_{f_{4}^{\prime}}^{j_{4}}\text{ }(x),\text{\ }%
\end{align}
where the index $k$ is the color one $k=1,2,3$ and the spinor indices are hidden to
simplify notation, $f$ indicates the flavor of the quarks. It should be
underlined  that the main different elements in this action with respect to
the massless QCD, are the presence of the two last terms and the change in the
sign of the Dirac Lagrangian. The last term is the added Nambu-Jona-Lasinio
like four quarks action.  \ The coefficients $\Lambda_{f_{1}\text{ }f_{2}f_{3}f_{4}}%
^{j_{1}j_{2}j_{3}j_{4}}$ are \ assumed to be consistent with all the
symmetries of QCD. For bookkeeping purposes, the conventions for the various
quantities are defined now as follows \
\begin{align}
F_{\mu\nu}^{a} &  =\partial_{\mu}A_{\nu}^{a}-\partial_{\nu}A_{\mu}^{a}-g\text{
}f^{abc}A_{\mu}^{b}A_{\nu}^{c},\nonumber\\
\Psi_{f}^{k}(x) &  \equiv\left(
\begin{array}
[c]{c}%
\Psi_{f}^{k,1}(x)\\
\Psi_{f}^{k,2}(x)\\
\Psi_{f}^{k,3}(x)\\
\Psi_{f}^{k,4}(x)
\end{array}
\right)  ,\text{ }\\
\Psi_{f}^{\dag k}(x) &  \equiv(\Psi_{f}^{k}(x))^{T\ast}\\
&  =\left(
\begin{array}
[c]{cccc}%
(\Psi_{f}^{k,1}(x))^{\ast} & (\Psi_{f}^{k,2}(x))^{\ast} & (\Psi_{f}%
^{k,3}(x))^{\ast} & (\Psi_{f}^{k,4}(x))^{\ast}%
\end{array}
\right)  ,
\end{align}
where $f$ $=1,...,6$ indicates the flavor index. The expressions for the
Dirac conjugate spinors and covariant derivatives are
\begin{align}
\overline{\Psi}_{f}^{j}\text{ }(x) &  =\Psi_{f}^{\dag k}(x)\gamma^{0},\text{
}\\
D_{\mu}^{ij} &  =\partial_{\mu}\delta^{ij}-i\text{ }g\text{ }A_{\mu}^{a}%
T_{a}^{ij},\ \overleftarrow{D}_{\mu}^{ij}=-\overleftarrow{\partial}_{\mu
}\delta^{ij}-i\text{ }g\text{ }A_{\mu}^{a}T_{a}^{ij},\\
D_{\mu}^{ab} &  =\partial_{\mu}\delta^{ab}-g\text{ }f^{abc}\text{ }A_{\mu}%
^{c},
\end{align}
in which the Dirac's matrices, $SU(3)$ generators and the metric tensor are
defined in this section in the conventions of reference  \cite{muta}, as
\begin{align}
\{\gamma^{\mu},\gamma^{\nu}\} &  =2g^{\mu\nu},\ \ \ \ [T_{a},T_{b}]=i\text{
}f^{abc}T_{c}.\text{ }\gamma^{0}=\beta,\text{ }\gamma^{j}=\beta\text{ }%
\alpha^{j},\text{ }j=1,2,3,\nonumber\\
g^{\mu\nu} &  \equiv\left(
\begin{array}
[c]{cccc}%
1 & 0 & 0 & 0\\
0 & -1 & 0 & 0\\
0 & 0 & -1 & 0\\
0 & 0 & 0 & 1
\end{array}
\right)  ,\text{ }\beta=\left(
\begin{array}
[c]{cc}%
I & 0\\
0 & -I
\end{array}
\right)  ,\text{ }\alpha^{j}\equiv\left(
\begin{array}
[c]{cc}%
0 & \sigma^{j}\\
\sigma^{j} & 0
\end{array}
\right)  ,\text{ }j=1,2,3,\nonumber\\
\sigma^{1} &  =\left(
\begin{array}
[c]{cc}%
0 & 1\\
1 & 0
\end{array}
\right)  ,\text{ }\sigma^{2}=\left(
\begin{array}
[c]{cc}%
0 & -i\\
i & -I
\end{array}
\right)  ,\text{ }\sigma^{3}=\left(
\begin{array}
[c]{cc}%
1 & 0\\
0 & -1
\end{array}
\right)  ,\text{ }I=\left(
\begin{array}
[c]{cc}%
1 & 0\\
0 & 1
\end{array}
\right)  .
\end{align}
\

Other precise definitions and relations for the coordinates are
\[
x\equiv x^{\mu}=(x^{0},\overrightarrow{x})=(x^{0},x^{1},x^{2},x^{3}),\text{
\ }x_{\mu}=g_{\mu\nu}x^{\nu},\text{ }x^{0}=t.
\]

As underlined before, a basic new element in the proposed action are \ the six vertices of the
form
\[
-\sum_{f}\varkappa_{f}\,\overline{\Psi}_{f}^{j}(x)\gamma_{\mu}%
\overleftarrow{D}^{ji\mu}\gamma_{\nu}D^{ik\nu}\Psi_{f}^{k}(x),
\]
where the six  coefficients $\varkappa_{f}$ will be called
\textquotedblleft condensate parameters\textquotedblright\ since they enter in
 similar `positions' of the  parameters appearing in the
\textquotedblleft motivating \textquotedblright\ non local vertex derived  in the
previous work \cite{16}. \ As before remarked, these terms play the relevant role in
the action of  allowing the validity  of the power counting
 renormalizability  of the model, even with the inclusion of the four
fermion terms. The $\varkappa_f$ are the only dimensional parameters in the \ theory.

The last new element with respect to the massless QCD in the action  is the
changed sign of the Dirac Lagrangian. The need of this change was discussed in
reference \cite{18}. This modification  leads to a free propagator which is expressed
as usual positive metric Dirac propagator of massive fermions plus a negative
metric massless propagators. The usual sign assignment determines that the
massive propagator shows negative metric. Since experiments  seem to indicate
that the massive quarks in QCD should be the ones which should be  physically
relevant within the model, the negative sign of the Dirac action was imposed.
However, \ it can be remarked that in reference  \cite{18} it was also argued
\ that such a change in the sign can be \ introduced in the same physical
action by a change of field and coordinates transformation. Thus, \ since the
same action is associated to the quantization procedure, the fixed negative
signs can be associated \ to a quantization of the same physical system but using alternative transformed
field variables and coordinates.

After  finding the inverse of the kernel associated with quadratic form in
the   quark fields, it follows that the quark propagator can be expressed as
the difference between a usual massive Dirac propagator and one also usual but
massless Dirac one, in the form
\begin{align}
S_{f}(p)  &  =\frac{1}{\gamma_{\nu}p^{\nu}-\varkappa_f\text{ }p^{2}}\nonumber\\
&  =\frac{1}{\gamma_{\nu}p^{\nu}}-(\frac{1}{\gamma_{\nu}p^{\nu}-m_{f}}).
\label{propa}
\end{align}
 The gluon free propagator is the usual one.  Then, both
propagators behaves as $\frac{1}{p2}$ and the maximal number of fields in a
Lagrangian term is 4,  therefore the model is power counting renormalizable.
The masses of the massive quarks become just the inverses of the six
condensate parameters \ $\varkappa_{f},$ $\ f=1,2,...5,6$. \ As it was
mentioned, the  massive propagator has the appropriate sign corresponding to
positive norm states. On another hand, the massless component has the sign
related with negative norm states. 

\begin{figure}[h]
\begin{centering}
\center
\includegraphics[scale=0.35]{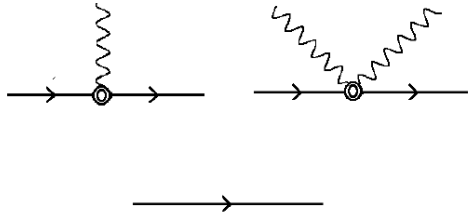}
\par\end{centering}
\caption{{}The two new types vertices, one pair for each flavor, determined by
the six new vertices.  The three legs one, at difference with the usual thee
legs vertex in usual QCD, is proportional to a product of two gamma matrices.
The two-gluon-two-quark four legs vertex is the local counterpart of the non
local vertex representing quark condensate effects obtained in reference. The line represents 
 the free quark propagator defined by (\ref{propa}).The
expectation associated with the approach presented in this work, is that the proposed
scheme can incorporate quark condensation effects in a way being able  in describing
a quark mass hierarchy.  }%
\label{vertices}%
\end{figure}

The new action terms also create two new vertices in the modified Feynman expansion as
depicted in figure \ref{vertices}. The line at the bottom represents the full quark propagator 
defined by (\ref{propa}).

For QCD, assumed to describe Nature, it is
currently interpreted that nor gluons or quarks show asymptotic states. Thus,
the negative metric of the massless free states seem not be a direct drawback
of the model. However, the fact that in very high energy processes, a
description based in massive quarks in short living asymptotic states, seems
to describe the experiences, suggests that an approach in which the massive
quarks have positive norms and massless do not appear thanks to radiative
corrections would be convenient. As remarked  before, this was the purpose of
the change in the sign of the Dirac action. For renormalizing the model, the
new two-gluons-two-quarks vertices should be included in the counterterm
action of the theory. But, as noted above, additional ones are  permitted by
the new more decreasing behavior of the quark propagator at large momenta.
Thus,  it follows that many of  the Lagrangian terms which define
the NJL models are allowed to be included as counterterms. \  Henceforth, the
mass generation properties embodied in such usually non-renormalizable
phenomenological theories, seem that can dynamically work now in the proposed
context. \

The  sum of fourth order terms in the quark fields should be a general
expression being invariant under the symmetries of QCD, to allow
the cancelation of the divergences. In the normal NJL theory, the usual Dirac
propagator, with its \textquotedblleft one over the momentum
modulus\textquotedblright\ behavior at large momenta, makes the model
non-renormalizable. Here, the one \textquotedblleft over modulus of the
momentum squared\textquotedblright\ behavior of the new quark propagators,
makes the Feynman expansion power counting renormalizable. This property can
be easily seen, by noting that the power counting rules for the proposed model are
identical to the ones in the simpler $\lambda \phi^4$ scalar field theory with the usual
scalar field propagator $\frac{1}{p^{2}}$ .

One important remark following  from the previous discussion should be added
here. Let us assume that a  renormalization procedure of  pure massless QCD is
being reconsidered. Then, the new two-gluons-two-quarks vertices, can be
identified as possible counterterms for this purpose, since they do not
destroy power counting renormalizability. Therefore, the\ idea comes to the
mind that the proposed model could result to be physically equivalent to the
quantized massless QCD.

In next section we simply expose an argue suggesting the possibility of the
theory to be also unitary. A closer investigation  of this property  is
expected to be considered in the extension of this work.

\section{Possible unitarity of the S matrix  in the Lee-Wick sense.}

The scattering among the physical particles of the proposed model should not
give rise to states in which unphysical particles exist as described by
propagating waves (unitarity). This property is illustrated in the picture at
the left: matrix elements of the scattering between a state defined by
incident physical states with states in which at least one non-physical
particle propagates should vanish.  Therefore, it should be checked whether or
not the first corrections to the quark propagator will allow the existence of
non physical propagating negative metric quark states in the proposed model.
\begin{figure}[h]
\begin{centering}
\center
\includegraphics[scale=0.50]{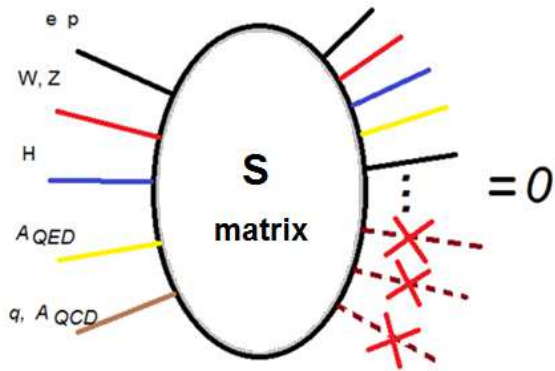}
\par\end{centering}
\label{mayor1}\caption{An illustration of the unitarity conditions for the S
matrix in the Lee-Wick sense.}%
\label{unitarity}%
\end{figure}
That is, showing poles in the momentum squared at positive values (mass
squared) for modes of negative metric which in the model are the massless
quark fields. The possibility for unitarity to be valid in the  theory is
strongly suggested by the existence of the so called Lee-Wick theories
\cite{leewick1,leewick2}. 
\begin{figure}[h]
\begin{centering}
\center
\includegraphics[scale=0.70]{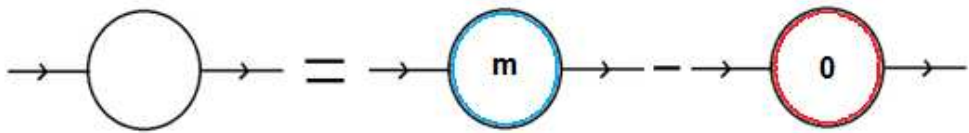}
\par\end{centering}
\caption{{}The picture illustrates the structure of the propagator in the
presented model. It is expressed as the sum of a positive metric massive propagator
plus negative metric massless one. This form leads to the
expectation \ that after evaluating the first radiative corrections \ the
propagator will not show poles corresponding to negative metric massless
states, at positive momentum squared values.  }%
\label{propagator}%
\end{figure}
\ In them, precisely helped by  the propagator being given as the
substraction of two propagators one of negative metric and another of positive
metric, as illustrated in figure  \ref{propagator},  these theories do not show
propagating negative metric states,  as a consequence of the  radiative
corrections \ to the propagators. Therefore, assumed that the \ model shows \ a
similar property, it can result to be unitary \ in the state space of the
\ physical particles in high energy scattering. The validity of this  property
is expected to be studied \ in the  extensions of the present work.

\section{A quark mass hierarchy from minimizing the vacuum energy?}

Finally, it  can be noted that the vacuum energy of the model, that is, minus
the effective action at vanishing mean fields, is a function of the six masses
$m_{f}$ $,$  $f=1,2,3...,6$ (the reciprocal of the condensate parameters
$\varkappa_{f}$). This quantity can be expected to show  all \textquotedblleft the
mass generating\textquotedblright\ properties  of the NJL model, which
are associated to the \textquotedblleft four fermions\textquotedblright\ vertices
entering the new diagram expansion. \  Therefore, if  we consider the
evaluation of the energy as a function of only two of those parameters,
\begin{figure}[h]
\begin{centering}
\center
\includegraphics[scale=0.70]{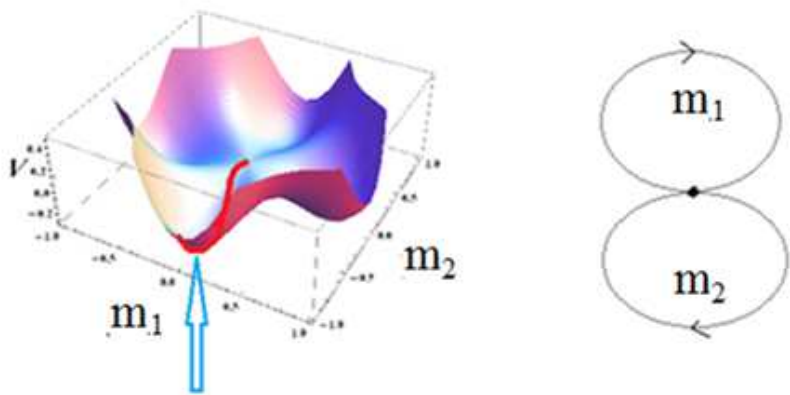}
\par\end{centering}
\caption{ The figure illustrates a  possible behavior which could be evaluated  from the model
for the effective potential  as a function of only two flavor mass parameters. The depicted  behavior
could be produced  if the  interference effects determined by the shown at right of the figure  two loop diagrams
(which can arise from the NJL vertices including closed loops for two different quarks) lead  to an enhance of the
energy for the configuration showing equal values of the quark masses. Then,  the minima of the potential
curve  could show  to be along the  axes, determining in this way a flavor symmetry breaking. The appearance of this behavior will be investigated in the extensions of the present work, after  the renormalization and the influence of all the kinds of  two loop vertices will be included.
     }%
\label{pot}%
\end{figure}
the fact that the fermion propagators are functions of these
two masses, and the existence of two loops \textquotedblleft interference
between flavors\textquotedblright\ effective potential diagrams, like the one
shown in the figure \ref{pot}, leads to the possibility that those terms make
the energy to rise when the two condensate takes similar values. Such an
effect will  lead to a flavor symmetry breaking in which one mass could
take a larger value with respect to the other. This could be the first
step in a hierarchy structure. In this connection it can be also remarked that
the two-gluon-two-quark diagrams directly break chiral invariance, thus the
appearance of masses is natural within  the model. This is am important
property which could allow the generation of a flavor symmetry breaking,
which requires the existence of a chiral non-invariance  as follows from
current algebra results \cite{current}.  These possibilities will be explored
in the extension of this work.

\section{Summary}

We reviewed a recently proposed improved version of the modified massless QCD
discussed in previous works. Motivated by it, it was proposed an
alternative to massless QCD including NJL action terms in a local and
renormalizable way. It is also underlined that the analysis done for
constructing the proposal suggests its equivalence with massless QCD. The new
terms determine masses for all the six quarks which are given by the
reciprocal of the new six flavor condensate couplings linked with each quark
type. The approach suggests a possibility to explain the quark mass hierarchy
as a dynamical flavor symmetry breaking. In it, the contributions of diagrams
showing two kinds of fermion lines might tend to rise the energy of the
configurations with equal values of the quark masses, making them more
energetic that ones in which a single quark mass parameter gets a finite
value.   It is interesting to remark that the occurrence of this flavor breaking,
in spite of  the known need of a chiral symmetry for the appearance of quark condensates which arose from the current algebra analysis, might be allowed by the fact  that included two-gluon-two-quark vertices directly break chiral invariance.  The considered framework seems appropriate to realize the so called
Democratic Symmetry Breaking properties of the mass hierarchy remarked by H.
Fritzsch \cite{fritzsch}. Finally,  it can be also imagined  that the appearance of six
different couplings in the theory, could be reduced  to only one by employing the
Zimmermann's reduction of the couplings approach \cite{zimmermann}. This possibility also
suggests a way for linking  the model with the SM  assuming  that the single coupling could be expected
to play the role of the Higgs field.  This property is also suggested by the known results
which show that the Top condensate models can be re-formulated  in a form being closer to the
SM.
It can be concluded  that the discussion  supports the starting idea of the
study about that massless QCD could generate an intense dimensional transmutation
effect. Its feasibility   will be investigated  in the extension
of this work. In ending, it must be remarked that in the gluodynamic limit
(which was not considered here) the appearance of Gaussian means over color
fields suggests the possibility of a first principles derivation of the linear
confining effects predicted by the stochastic vacuum models of QCD
\cite{doetsch}.

\section*{Acknowledgments}
  I would like to  express  my  acknowledgements to the Department of Physics of the New York University by the kind hospitality during a recent short visit in which this work was exposed there, and  the helpful remarks received during the stay from  M. Porrati, G. Gabadadze, A. Reban, J. Lowenstein, D. Zwanziger and A. Sirlin.  I should also express my gratitude  by the conversations about  this work sustained with  A. Gonz\'alez, A. Tureanu, M. Chaichian and A. Klemm. The support granted by the N-35 OEA Network  of the ICTP is also greatly appreciated.

\end{document}